\documentclass[amsmath,amssymb,prl,twocolumn,superscriptaddress]{revtex4-2}

\usepackage[utf8]{inputenc}
\usepackage{graphicx}
\usepackage{dcolumn}
\usepackage{bm}
\usepackage{siunitx}
\usepackage[version=4]{mhchem}
\usepackage{natbib}
\usepackage{float}

\usepackage[shortlabels]{enumitem}

\usepackage{xcolor}

\usepackage{mathtools}
\usepackage{braket}
\usepackage{amsmath}%
\usepackage{MnSymbol}%
\usepackage{wasysym}%
\usepackage{cancel}
\usepackage{physics}
\usepackage{caption}
\usepackage{subcaption}
\usepackage[english]{babel}
\usepackage[T1]{fontenc}
\usepackage{placeins}
\usepackage[normalem]{ulem}

\usepackage{hyperref}

\makeatletter
\def\maketitle{
\@author@finish
\title@column\titleblock@produce
\suppressfloats[t]}
\makeatother
\newcommand{\beginsupplement}{
    \onecolumngrid
    \setcounter{table}{0}
    \renewcommand{\thetable}{S\arabic{table}}
    \setcounter{figure}{0}
    \renewcommand{\thefigure}{S\arabic{figure}}
    \setcounter{equation}{0}
    \renewcommand{\theequation}{S\arabic{equation}}
}

\begin{document}

\title{Comparison of Origins of Re-Entrant Supercurrents at High In-Plane Magnetic Fields in Planar InAs-Al Josephson Junctions}

\author{S.R. Mudi}
\affiliation{Department of Physics and Astronomy, University of Pittsburgh, PA 15260, USA}

\author{S. Anupam}
\affiliation{JARA-FIT Institute for Quantum Information, Forschungszentrum J$\ddot{u}$lich GmbH, Campus-Boulevard 79, 52074 Aachen, Germany}

\author{V. Mourik}
\affiliation{JARA-FIT Institute for Quantum Information, Forschungszentrum J$\ddot{u}$lich GmbH, Campus-Boulevard 79, 52074 Aachen, Germany}

\author{S.M. Frolov}
 \email{frolovsm@pitt.edu}
\affiliation{Department of Physics and Astronomy, University of Pittsburgh, PA 15260, USA}

\date{\today}

\begin{abstract}
Hybrid superconductor-semiconductor systems with large spin-orbit coupling are important platforms for realizing topological or triplet superconductivity. Planar Josephson junctions made using these materials are predicted to enter the topological state by tuning the phase difference between the two superconductors from 0 to $\pi$. The 0-$\pi$ transition can be driven by magnetic field through Zeeman splitting of subbands in the semiconductor. It is expected to manifest as a node, or a re-entrance, in the critical current. Here we present re-entrant switching currents from several InAs/Al planar Josephson junctions in high in-plane magnetic fields. We find that re-entrances in some devices conform with expected signatures for topological or 0-$\pi$ transitions. However, we show that the data can also be explained in terms of mode interference in the junction in the presence of disorder. We also present simulations of supercurrent interference under in-plane fields that can reproduce re-entrances due to corrugated weak link without invoking the Zeeman effect or topology.
\end{abstract}

\maketitle

\captionsetup{justification=raggedright,singlelinecheck=false}

\subsection{Introduction}

Josephson junctions with different current-phase relations (CPRs) are important for quantum computing applications \cite{feofanov2010implementation,gladchenko2009superconducting,larsen2020parity,frattini20173,reza2025predicted}. In this regard, $\pi$-junctions are interesting because the ground state phase difference changes from $0$ to $\pi$ leading to an inverted CPR $I_c \mathrm{sin} \phi \rightarrow I_c \mathrm{sin}(\phi + \pi) = -I_c \mathrm{sin} \phi$.  Such $\pi$ junctions have been successfully demonstrated using exchange interaction in ferromagnetic barriers \cite{ryazanov2001coupling,frolov2004measurement} and gate-tunable quantum dot-based junctions \cite{cleuziou2006carbon,van2006supercurrent}. Controllable $\pi$ junctions can also be achieved by modulating the energy distribution of the quasiparticles in the normal metal \cite{baselmans2002direct}. Another approach is to change the properties of the superconducting leads, for example, using d-wave superconductors which have an unconventional order parameter \cite{wollman1993experimental}. In these experiments, the 0-$\pi$ transition was controlled by changing the junction length, temperature, or voltages on the electrostatic gates or junction leads.

\par $0-\pi$ transitions are of interest in the context of Majorana Zero Modes (MZMs), which could lead to a version of fault-tolerant quantum computation \cite{kitaev2001unpaired}. These modes have been predicted to appear in materials with large spin-orbit coupling (SOC) when combined with conventional s-wave superconductors \cite{PhysRevLett.100.096407, lutchyn2010majorana}. Tuning the chemical potential and magnetic field to prescribed values can drive the hybrid device into the topological state. A 0-$\pi$ transition can also lead to MZM formation~\cite{pientka2017topological,dartiailh2021phase}. In this case, a re-entrance of supercurrent is observed at finite magnetic field, oriented in the junction plane to minimize orbital depairing \cite{crosser2008nonequilibrium}. Alternatively, a superconductor-semiconductor planar Josephson junction embedded in a loop can also host MZMs when an external phase bias of $\pi$ is applied \cite{pientka2017topological}. This idea was experimentally explored in Refs.~\cite{fornieri2019evidence, ren2019topological} in the presence of a substantial in-plane field. These measurements were reporting tunneling spectroscopy which is prone to false positive MZM signals due to disorder \cite{lee2014spin,chen2019ubiquitous}.

Supercurrent re-entrances were reported in several planar junctions such as the 2D topological insulator HgTe coupled to Al and Nb \cite{hart2017controlled}, InSb/NbTiN \cite{ke2019ballistic} and InAs/Al \cite{dartiailh2021phase}. In all these reports, gate and in-plane magnetic field tunable nodes in the switching current were observed and were attributed to finite momentum pairing. In the 3D topological insulator Bi$_2$Se$_3$ coupled to NbTi/NbTiN \cite{chen2018finite}, anomalous interference patterns were attributed to finite Cooper pair momentum. 

\begin{figure*}[ht]
    \centering
    \includegraphics[width=0.9\textwidth]{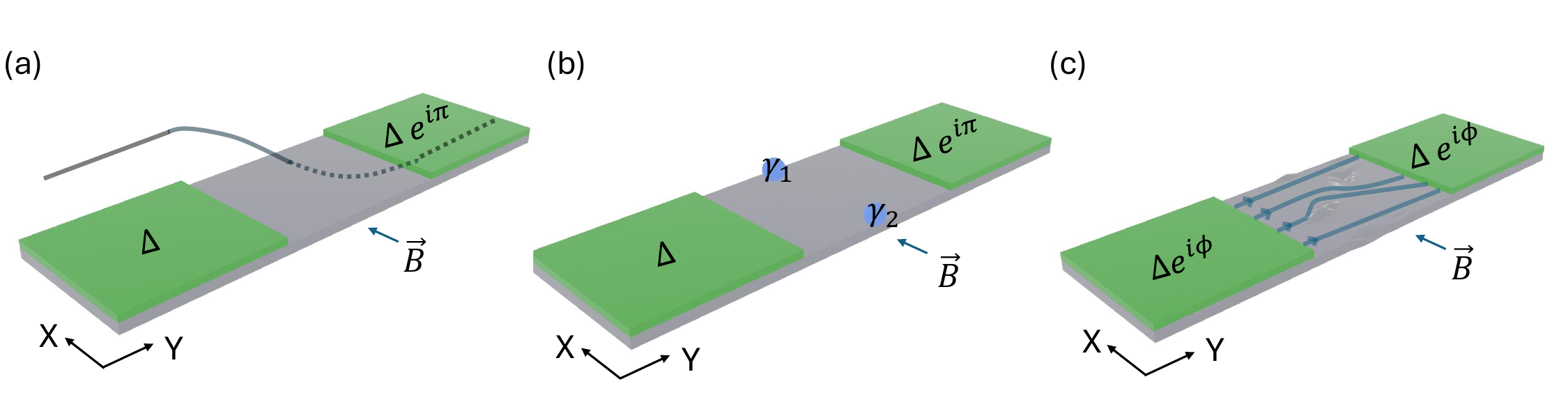}
    \caption {Schematic for the three processes leading to re-entrant supercurrents : (a) A trivial Zeeman $\pi$ junction with an antisymmetric order parameter denoted by the black curve, (b) A planar junction with spin-orbit coupling under an in-plane magnetic field leads to a $\pi$ phase difference and the appearance of Majorana Zero Modes at the ends of the junction, (c) A disordered junction which can make the electrons deviate from straight line paths and interfere with each other. The gray and green colors represent the superconductor and the quantum well respectively.}
    
    \label{Fig1_Reentrant_Mechanisms}
\end{figure*}

\par Studies in other hybrid systems have also reported re-entrant supercurrents, not due to topology or 0-$\pi$ transitions. In InSb/NbTiN nanowire devices \cite{zuo2017supercurrent}, several supercurrent nodes with a strong dependence on the chemical potential were observed. In that work, the interplay of disorder along with orbital, Zeeman and SOC was analyzed in the few mode regime to explain the re-entrances. Supercurrent interference due to orbital effects was used to explain re-entrances in InAs/Nb nanowires \cite{gharavi2014josephson}. Graphene/NbSe$_2$ junctions, where no topological effects are expected, also reported several nodes in the data at high in-plane magnetic fields \cite{dvir2021planar}. There, the re-entrances were attributed to ripples in graphene, which caused the supercurrent not to be truly two-dimensional, but rather have a three-dimensional character. Experiments on systems such as tri-layer graphene \cite{cao2021pauli} and single crystal UTe$_2$ \cite{aoki2022unconventional} also showed re-entrances at high fields. The findings were attributed to p-wave or 'non-singlet' superconductivity, since high magnetic fields favor the alignment of spins and can, in principle, stabilize a spin-triplet state. These works are not based on junctions and induced superconductivity and therefore our analysis does not directly apply to them.

In this report, we investigate re-entrant supercurrents in planar Josephson junctions based on InAs two-dimensional electron gases (2DEGs). We analyze our findings in the context of both the exotic interpretations and disorder. We present re-entrant supercurrent data from several planar junctions, including those where re-entrant supercurrents do not resemble $0-\pi$ transitions. We observe a wide variety of supercurrent interference patterns that change with the top gate voltage and magnetic field. We explore the similarities and differences in these patters across several devices of different geometries. Some devices show a single node at a fixed in-plane field, irrespective of the out-of-plane field applied, and the node moves to higher in-plane field as the gate voltage is decreased. Other devices show re-entrances whose in-plane field position changes with the out-of-plane field and the gate voltage and does not follow a particular trend, not reported previously in this system. We attempt to emulate some of our data using a numerical model that allows for deviations from a perfectly flat junction. Our work highlights the importance of considering trivial effects when interpreting data in the context of topology \cite{frolov2023smoking}.

\begin{figure*}[ht]
    \centering
    \includegraphics[width=0.9\textwidth]{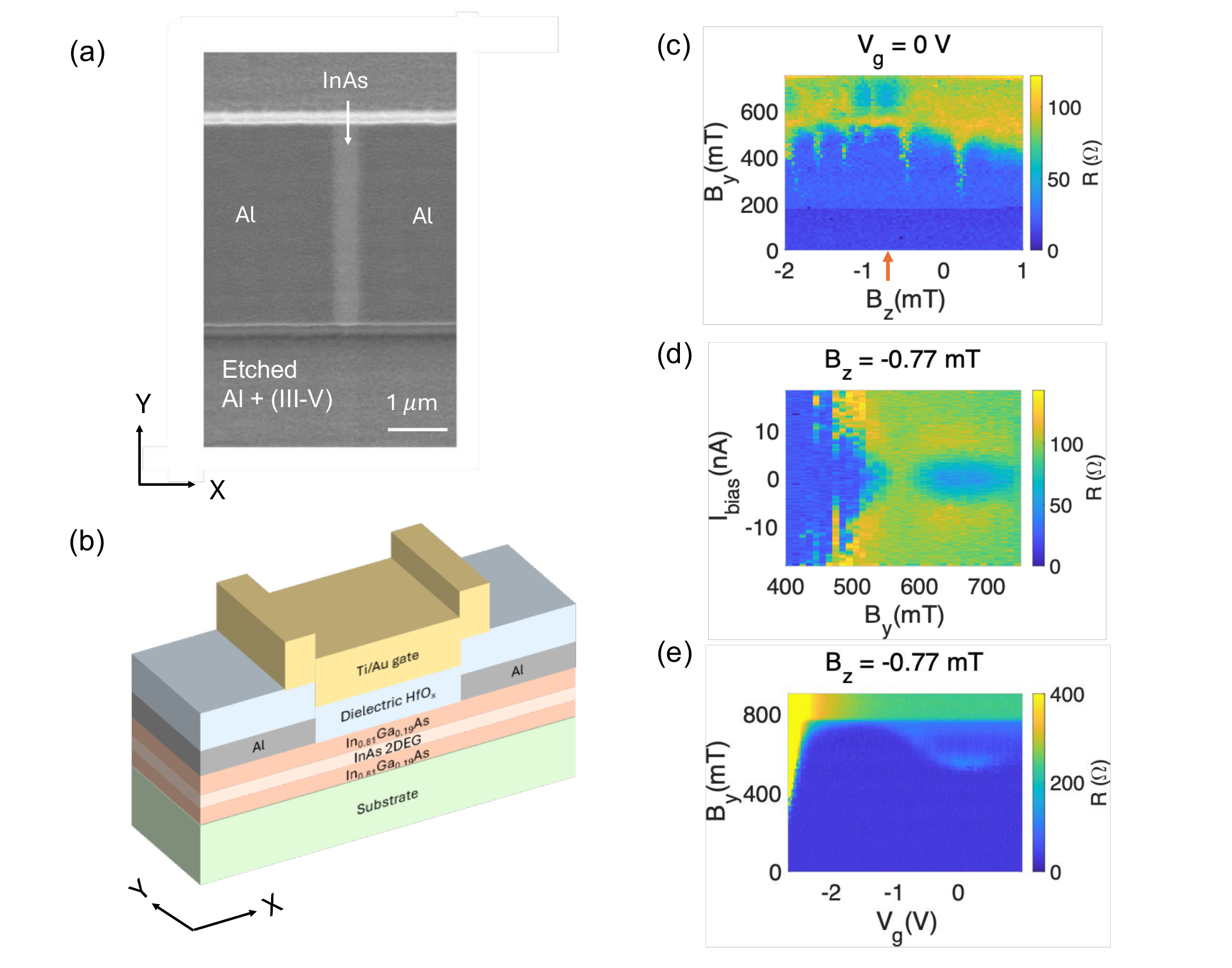}
    \caption{Supercurrent interference data for Device 1 : (a) SEM Image of the planar InAs/Al Josephson junction. Image was taken before deposition of the dielectric and top gate. (b) Schematic of the device and heterostructure stack, with $x$ and $y$ directions indicated. (c) Differential resistance as a function the in-plane ($B_y$) and out-of plane ($B_z$) fields at DC bias = 0, (d) Differential resistance as a function of the dc current bias and $B_y$ at $B_z = -0.77$ mT (corresponds to the orange arrow in (c)), (e) Differential resistance as a function of $B_y$ and top gate voltage $V_g$ at $B_z = -0.77$ mT. A lock-in excitation of 2 nA at 43 Hz was used for (c), (d) and (e).}
    \label{Reentrant_Fig2_UniformJunctionGateDependence}
\end{figure*}

\subsection{Experimental Details}
We fabricate InAs/Al Josephson junctions using InAs near-surface quantum wells grown on InP substrates and sandwiched between InGaAs barrier layers (grown by Ido Levy and Javad Shabani at NYU). The Al thickness is approximately 15 nm for the devices presented here. Because several devices are fabricated on the same chip, the Al and InAs are first etched around the devices to isolate the mesas. A thin strip of Al is then etched to create a break in the Al layer to form the junction. This is followed by the deposition of dielectric HfO$_x$ at 80 $^o$ C, followed by the deposition of Ti/Au top gates. Further details of the recipes used for device fabrication are provided in the Supplementary Information. All measurements for this study were performed in a two-terminal configuration in dilution refrigerators with base temperatures ranging from 12-65 mK. The setups are equipped with vector magnets (9-1-1 T or 6-4 T). 

The measurements are performed in a two-terminal configuration where a dc current bias is applied to the device along with an ac current excitation of 1-10 nA at a frequency of 37-77 Hz using a lockin and the averaged voltage across the device is measured. The exact lockin frequency and excitation used for each dataset is noted in the respective figure captions. The dc current bias is set to zero for all in-plane versus out-of-plane field scans unless otherwise noted. We note that the lockin excitation used can affect the supercurrent reentrance pattern measured. For example, if the lockin excitation is smaller than the supercurrent value at the node, the node will not show up in the data. This would be most relevant for the lower in-plane field values where the switching current value is still high, but may already display non-monotonous decay or nodes and partial revivals. On the other hand, at higher in-plane magnetic fields, where the supercurrent is small and the junction's transitions from the superconducting to the resistive state are not sharp, the lockin excitation may also not be able to capture the changes in the switching current values accurately, as such subtle changes may manifest themselves as slight changes in the integrated resistance. For all figures, we define the z-axis as the out-of-plane direction to the chip, the x-axis as the direction of current flow and the y-axis as the direction perpendicular to the current flow. Devices were fabricated in two orthogonal orientations on the chip, with a possible offset angle between the device and the coordinate system defined by the vector magnet used in the experiment due to fabrication and small rotations during the mounting of the chip onto the sample holder. $B_{x,y,z}$ therefore nominally point along the directions defined above.

Since we perform two terminal measurements, the raw differential resistance is not zero in the supercurrent state. For each plot, we subtract the minimum resistance for that plot to make the superconducting state correspond to zero resistance. Although the minimum resistance measured for each device is nominally the same (within $\approx$ 200 $\Omega$), there maybe slight differences in the dc lines, the contact resistance due to wire bonding, charge jumps, etc. There was a difference between lock-in and DC values of the measured resistance. This is caused by the chosen lock-in frequency in our differential measurements being close to the low-pass filter's cut-off in the set-up, leading to a lowering of the resistance of order 10\%. To address this, for the lock-in based measurements, we scale the differential resistance by 4.2 $k\Omega$, the known series resistance of the setup, divided by the minimum resistance to estimate the actual value of the resistance at each data point. This data processing step affects the magnitude of the signal plotted, but not the recorded patterns in the data. Therefore, it does not affect the analysis, which is primarily based on supercurrent/no supercurrent classification of 2D colormaps.

\subsection{Figure 1 description}

We summarize the various physical phenomena that could lead to supercurrent re-entrances in Fig. \ref{Fig1_Reentrant_Mechanisms}. In Fig. \ref{Fig1_Reentrant_Mechanisms}(a), we represent a Josephson $\pi$ junction where a magnetic field lifts the degeneracy of the two spin bands \cite{demler1997superconducting}. The Fermi surfaces of the two bands shift giving the center of mass of the Cooper pairs a finite momentum, which results in an oscillating order parameter. By tuning the magnetic field, the oscillation period and hence the order parameter overlap in the weak link can be changed. When the Zeeman splitting is large enough, the ground state energy is minimized for a phase difference of $\pi$. At the $0-\pi$ transition point, a minimum in the critical current is reached. The critical current grows away from the transition point, exhibiting a re-entrant behavior.

\par When the semiconductor weak link has spin-orbit coupling, an in-plane magnetic field can theoretically drive the junction into the topological state with MZMs appearing at the ends of the junction, as shown in Fig. \ref{Fig1_Reentrant_Mechanisms}(b). In this version of the effect, the phase shift of $\pi$ would provide the missing ingredient to complete the topological transition, assuming that the weak link can be viewed as a quasi-one-dimensional longitudinal mode. A re-entrant supercurrent then should coincide with the closing and re-opening of the induced superconducting gap \cite{dartiailh2021phase}.

\par Re-entrances can also arise due to supercurrent interference in the presence of disorder. This scenario is depicted in Fig. \ref{Fig1_Reentrant_Mechanisms}(c). A planar junction has multiple transverse one-dimensional modes due to its finite width. In the absence of disorder, the modes align perfectly parallel to each other in the plane, making the effective junction thickness equal to zero. This implies that an in-plane field will result in zero flux through the effective cross-sectional area, i.e., no supercurrent interference due to mode mixing.  

Planar junctions fabricated using quantum wells grown by molecular beam epitaxy are considered to have a smooth surface due to the high quality of growth. However, the presence of disorder can alter this picture by shifting the vertical position of the 2DEG locally. This can result from doping variations, surface corrugations during growth, effects from nanofabrication on the 2DEG itself as well as due to screening in nearby metals and Meissner effect in the superconducting contacts, the effects from the granularity of the metals around the junction. 

For example, if the mean thickness of a two-dimensional electron gas (2DEG)  varies by 1 nm for a junction width of 2 $\mu m$, then a magnetic field of 1 T will enclose one flux quantum in the junction when applied in-plane. This can lead to supercurrent interference and show up as a minimum in the critical current, similar to nodes in Fraunhofer patterns for out-of-plane magnetic fields \cite{tinkham1996introduction}. In a less disordered junction, the interference minimum will be pushed to higher magnetic fields, while in a more disordered junction, re-entrances can happen at lower fields~\cite{hart2017controlled}. 

\subsection{Figure 2 description}

In Figure \ref{Reentrant_Fig2_UniformJunctionGateDependence}, we present supercurrent re-entrance data for Device 1. Fig. \ref{Reentrant_Fig2_UniformJunctionGateDependence}(a) shows an SEM image of the device taken before the deposition of a top gate, which is the case for all images in this paper. A common approach to studying re-entrances is to vary both the in-plane and out-of-plane magnetic fields \cite{keijzers2012josephson}. The out-of-plane fields are used to record the standard diffraction patterns where critical current minima appear at field values of the order of a flux quantum over the junction area, which is a much smaller field scale than the re-entrances for in-plane fields. The assumption is that the diffraction patterns remain symmetric and re-entrance can be identified by a minimum in the center of the pattern. The diffraction pattern often shifts to higher z-fields, either due to geometrical considerations such as the orientation of the chip with respect to the vector magnet, trapped vortices, bending of the PCB at high fields, etc. (see Supplementary Information Fig. \ref{ShiftedFraunhofer} for an example). Therefore, for every applied in-plane field value, the z-field range scanned was adjusted in real time to keep the central and first few lobes of the diffraction pattern in the field of view. The misalignment angle between in-plane and out-of-plane fields is approximately $\pm 2^{\circ}$ for devices 1-5 and $\pm 5^{\circ}$ for devices 6-7. This adjustment was done for all in-plane versus out-of-plane field plots for all devices. The value used for adjustment differed slightly between devices for a given in-plane vs out-of-plane field direction.

In Fig. \ref{Reentrant_Fig2_UniformJunctionGateDependence}(c), the x-axis represents the effective z-field values after the real time adjustment of the z-field based on the applied $B_y$. The differential resistance is obtained at zero DC bias plus a lockin excitation. An increase in the resistance corresponds to a minimum in the supercurrent. The dark blue regions denote the junction's low resistance state. The jump in resistance in the superconducting state in Fig. \ref{Reentrant_Fig2_UniformJunctionGateDependence}(c) is possibly due to a charge jump.

As the in-plane field increases we observe a strong suppression of supercurrent around $B_x$=550 mT for all out-of-plane ($z$) fields. Supercurrent re-appears beyond $B_x$=580 mT for certain $z$-fields, indicating a re-entrance of the supercurrent. The re-entrances can also be visualized by fixing the z-field where a minimum in the supercurrent is obtained and sweeping both the current bias and the in-plane field. This is shown in Fig. \ref{Reentrant_Fig2_UniformJunctionGateDependence}(d) where the z-field is fixed at -0.77 mT (orange arrow in Fig. \ref{Reentrant_Fig2_UniformJunctionGateDependence}(c)). As the in-plane field increases, the supercurrent decreases and goes to zero at $B_x$ = 550 mT, stays zero until $B_x$ = 580 mT, and then reappears again, persisting until $B_x$ = 750 mT. 

\par In Fig.~\ref{Reentrant_Fig2_UniformJunctionGateDependence}(e), we show the top gate voltage dependence of the node at $B_z$ = -0.77 mT. The node appears in the data between -1 V$\le V_g \le$ 0.5 V, with the position of the node moving to higher $B_x$ fields as the gate voltage is reduced. The node's behavior is qualitatively similar in appearance to the data presented in Refs.~\cite{dartiailh2021phase,hart2017controlled}, where these are attributed to topological origins. The position of the node is robust under multiple gate sweeps over a gate voltage range of order volts for this device. 

To demonstrate the re-entrances more clearly, the colorbar limits in Fig.~\ref{Reentrant_Fig2_UniformJunctionGateDependence}(e) were adjusted. See supplementary materials for the raw data (Fig. \ref{Fig2d_full}). Beyond the field of approximately 750 mT, Al transitions into the non-superconducting state. The signal is weak beyond the re-entrance point, meaning that the switch in the current-voltage characteristics that marks the boundary of the blue region is smooth, and the switching point likely does not represent the true critical current of the junction. At other z-field values, the signal may be even weaker and beyond the detection threshold. In this regime, the true shape of the Josephson critical current diffraction pattern cannot be extracted from the transport measurements.

\subsection{Figure 3 and 4 description}

\begin{figure*}[ht]
    \centering
    \includegraphics[width=0.93\textwidth]{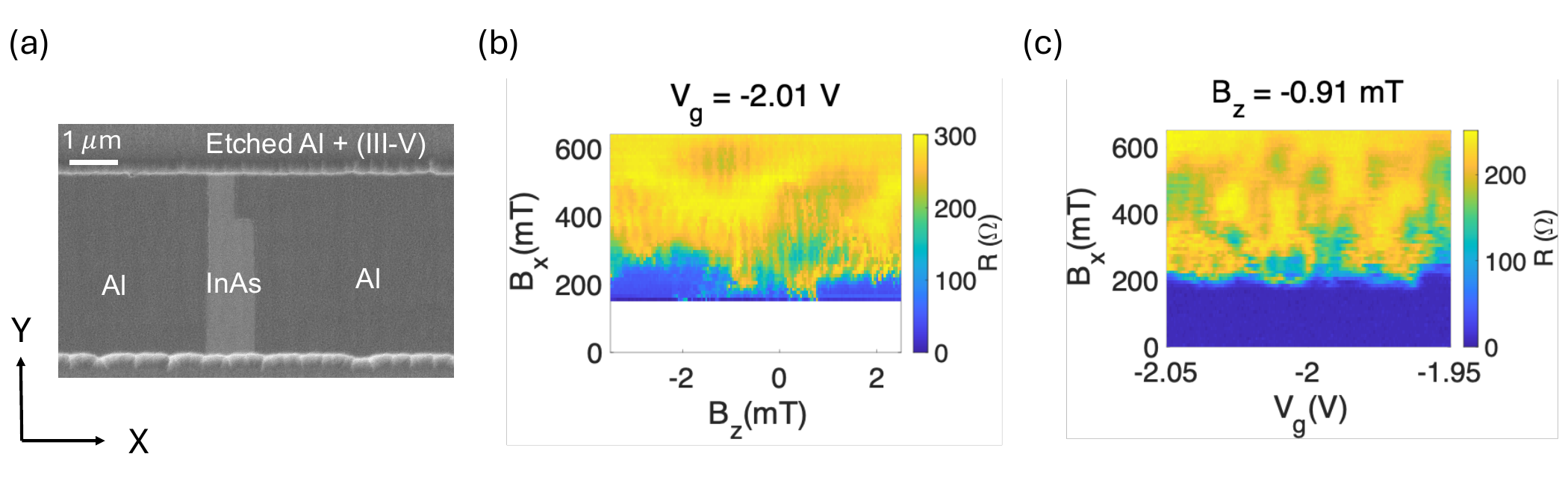}
    \caption {Supercurrent interference data for Device 2: (a) SEM of the junction with two lengths $\approx$ 550 nm and 988 nm, (b) Differential resistance as a function of the in-plane and out-of-plane magnetic fields $B_x$ and $B_z$ at $V_g$ = -2.01 V, (c) Differential resistance as a function of $B_x$ and top gate voltage $V_g$ at $B_z$ = -0.91 mT. A lockin excitation of 1 nA at 49 Hz was used for (b) and (c).}
    
    \label{Reentrant_Fig3_TwoWidthJunction}
\end{figure*}

\begin{figure*}[ht]
    \centering
    \includegraphics[width=0.9\textwidth]{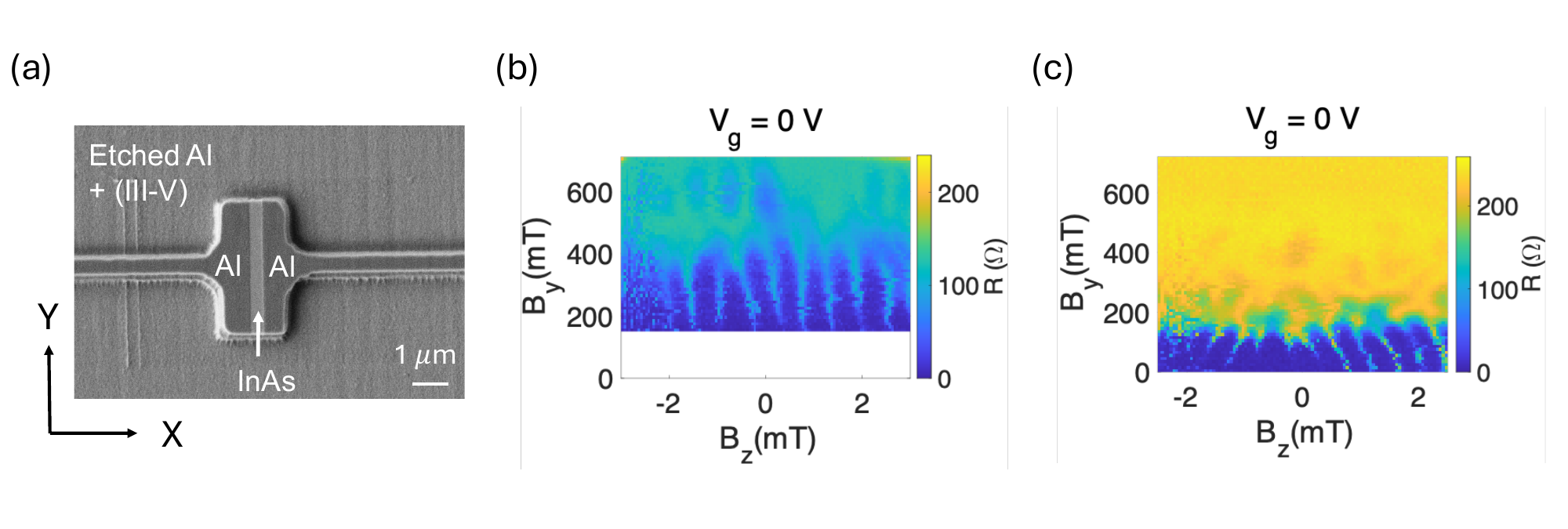}
    \caption {(a) SEM image of Device 4, (b) Supercurrent interference pattern for Device 4, and (b) Supercurrent interference pattern for Device 5 which looks similar to Device 4 (Supplementary Information Fig. \ref{Reentrant_Device4}(a). A lockin excitation of 1 nA at 43 Hz was used for both (b) and (c).}
    
    \label{Reentrant_Fig4_PatternVariety}
\end{figure*}

Next, we present data from several samples fabricated using the same steps as Device 1 aside from geometric variations. Data for Device 2 are shown in Figs. \ref{Reentrant_Fig3_TwoWidthJunction}(b) and (c). We interpret the brightest yellow color as regions without detectable supercurrent signal, while darker colors (orange, green and blue) correspond to some supercurrents. Since Device 2 is wider than Device 1, we observe on the order of 25 diffraction pattern nodes in panel (b). We also see that the node positions shift back and forth in $B_z$ even after the subtraction of a linear offset due to a small angle between the chip plane and the magnet z-axis. At higher in-plane fields above 400 mT, the stronger supercurrent signals appear at offset $B_z$ values. The diffraction patterns lack clear symmetry in $B_z$. Note that the in-plane field is applied along the current flow for measurements on Device 2.

Figure \ref{Reentrant_Fig3_TwoWidthJunction}(c) shows the gate dependence at $B_z$ = -0.91 mT. At higher in-plane fields above 200 mT, the superconducting regions are strongly tunable by the gate voltage. Some of this is due to the larger area of the etched weak link compared to Device 1. Even accounting for stronger gate coupling, the pattern appears rich and lacking particular structure. Fixing either gate voltage or an offset $B_z$ value, it should be possible to produce a variety of re-entrant dependences of the switching current on the in-plane field.

In Figs.~\ref{Reentrant_Fig4_PatternVariety}, we show interference patterns from two additional devices of similar geometry. The in-plane field for both devices was applied approximately perpendicular to the current flow direction. The interference patterns here combine the features illustrated with data from Devices 1 and 2. In Device 4 (Fig. \ref{Reentrant_Fig4_PatternVariety}(b)), the re-entrances at high fields all appear at $B_x$ = 540 mT, similar to Device 1, whereas there are several re-entrances that appear for Device 5 (Fig. \ref{Reentrant_Fig4_PatternVariety} (c)) at different $B_y$ values based on the z-field value.

Some of the devices have a step in the junction length (see supplementary information for SEM images). This is in order to create a 0-$\pi$ junction and observe a diffraction pattern minimum when the longer part enters the $\pi$ state \cite{frolov2006josephson}. The step may be considered a defect that is causing additional interferences, however an interference pattern from a single defect is expected to be organized. The patterns presented in this Figure are consistent with multiple defects. For some of the junctions, etching resulted in rougher boundaries which is one identifiable source of disorder. More devices are shown in the supplementary information. 

\subsection{Figure 5 description}

\begin{figure*}[ht]
    \centering
    \includegraphics[width=0.9\textwidth]{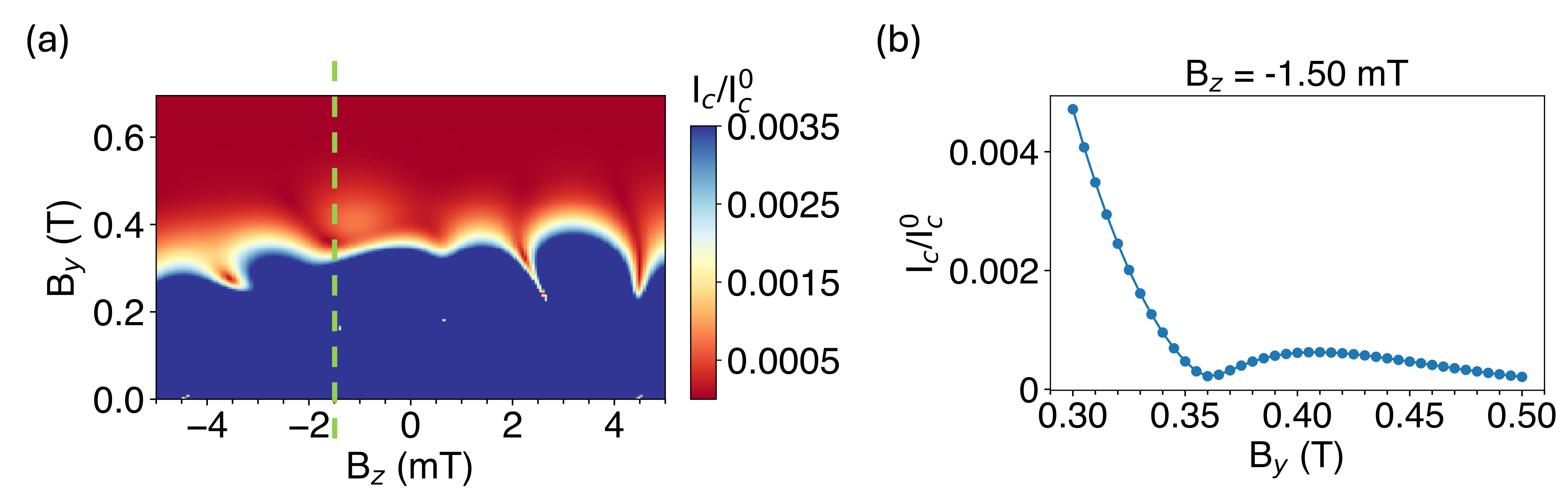}
    \caption {Results of numerical simulations : (a) Normalized critical current as a function of the out-of-plane and in-plane magnetic fields, (b) Linecut showing a minimum in the critical current at $B_z$ = -1.5 mT (corresponds to the green dashed line in (a)).}
    
    \label{Reentrant_Fig5_RandDisorder}
\end{figure*}

Fig. \ref{Reentrant_Fig5_RandDisorder} presents semiclassical numerical simulations of a supercurrent interference pattern in a disordered Josephson junction under applied in-plane and out-of-plane magnetic fields (separate manuscript in preparation). Here, disorder is modeled as a modulation of the junction surface, and consequently of the supercurrent. This may represent physical defects such as step edges and wrinkles, or spatial variations in the potential landscape across the junction, arising from sources such as interfacial charge traps, doping inhomogeneities, nonuniform contacting, or fabrication-related imperfections. An in-plane magnetic field applied to such a nonuniform surface, over which the supercurrent flows, can lead to trapped flux that would not occur in an atomically flat junction.

In the model, the surface profile is generated by specifying a disorder density, strength, and spatial extent. The disorder density and strength are sampled from independent normal distributions, and a Gaussian filter is applied to control the spatial spread of the disorder centers. By tuning these parameters, the model produces a broad range of interference patterns, including features that can resemble previously reported signatures attributed to nontrivial mechanisms. Cherry-picking through this parameter space allows us to identify a pattern exhibiting a re-entrance at approximately \SI{350}{mT}, as shown in Fig. \ref{Reentrant_Fig5_RandDisorder}(a). Fig. \ref{Reentrant_Fig5_RandDisorder}(b) shows a line cut taken at a fixed out-of-plane magnetic field, highlighting the re-entrance with greater clarity.

Because the Fraunhofer simulations do not incorporate supercurrent decay in magnetic field, an ad hoc envelope is introduced, with distinct field dependence along different magnetic field directions. Additional adjustments include truncation of the color scale to emulate the finite detection range of the lock-in amplifier, arising from its relatively large excitation amplitude compared to the switching current. These modifications produce an effective truncation and “fisheye”-like distortion in the interference pattern, while suppressing residual supercurrent revivals and low-field substructure. In experiments, such re-entrance features may be obscured in differential resistance measurements, for example due to offsets. Additional re-entrances may also occur at other combinations of in-plane and out-of-plane fields but remain unobserved due to experimental settings or measurement offsets.

The simulations can be further extended by introducing sharp disorder features near the contact regions or along the junction edges, mimicking contact disorder or etching-induced defects in real devices. This can give rise to an in-plane angular dependence of the supercurrent oscillations. For example, an elongated, buckled region - where the junction plane is locally displaced upward or downward, with a small cross-sectional area perpendicular to the current flow compared to along it — could account for some observations in Device 1. When an in-plane magnetic field is applied perpendicular to the current flow direction, such a geometry can produce a single re-entrance arising from supercurrent interference.

\subsection{Conclusions}

\par In conclusion, we present data showing re-entrant supercurrents in planar InAs/Al Josephson junctions, a system predicted to host topological states. We study both the magnetic field and gate dependence of supercurrents and compare our results with signatures expected for topological transitions. Some of our devices show a single, robust re-entrance as a function of the chemical potential and magnetic field, and the magnetic field at which the feature is observed agrees with back of the envelope calculations for a 0-$\pi$ junction or a topological transition. Others junctions show several re-entrances in a narrow gate range that change with fast gate sweeps. 

Some junctions show features that fall in between these two cases, with some semblance of organized interference patterns at high magnetic fields, yet no clear symmetry. Supercurrent signals typically weaken at higher in-plane fields which can result in some of the features losing visibility, and sometimes leaving those consistent with a clean and simple re-entrance, the only observed feature due to its central location in the diffraction pattern where the signal is the strongest. 

While this is not the focus of this paper, we also notice that the effective junction length extracted from the out-of-plane periodicity of the diffraction patterns is of order 1-2 microns, for a lithographic width of 200-400 nm. This size would be the transverse dimension for a topological nanowire defined within the junction, but the mode quantization in such a wide junction would be of order $\mu eV$ putting the few-subband regime needed for realizing MZM firmly out of reach.

Taken as a whole, our results are consistent with a disorder-based interference picture, as illustrated by the numerical simulations. The presence of disorder results in an effective finite junction thickness for a nominally two-dimensional system, which can lead to nodes of destructive supercurrent interference in an in-plane field. Such considerations are important in the context of topological and triplet superconductivity because re-entrances have been reported in planar junctions and quasi two dimensional systems in this context. Reporting of more extensive datasets should make it possible to estimate what role disorder plays in shaping the interference patterns in a given set of samples \cite{frolov2023smoking}.

\section{Acknowledgements}
We thank I. Levy and J. Shabani from NYU for growing the quantum well samples used in this study.

\section{Funding}

Experimental work is supported by ONR.

\section{Data and Code Availability}
The experimental datasets and the code for plotting the figures can be found \href{https://doi.org/10.5281/zenodo.19180406}{here}.

The code for the theory plot can be found \href{https://gitlab.com/squad-lab/theory/fraunhofer/-/tree/main/cases/Pittsburgh?ref_type=heads}{here}.

\section{Duration and Volume of Study}
The study was conducted between May 2023-May 2025. During this time, 91 devices were measured and about 2500 data sets were collected.

\section{Author Contributions}
S.R.M fabricated the devices and performed measurements. S.M.F. and S.R.M. analyzed the data. S.A. and V.M. performed simulations. All authors contributed to the writing of the manuscript.

\bibliography{main.bib}
\bibliographystyle{ieeetr}

\clearpage
\title{Supplementary Information: Origins of Re-Entrant Supercurrents in Planar Josephson Junctions}
\maketitle
\beginsupplement

\begin{table*}[t]
\begin{ruledtabular}
\begin{tabular}{p{1.5cm}p{1.5cm}p{1.5cm}p{1.5cm}}
Wafer& Chip Number& Device Number on Chip& Device Name in the text\\
\hline
JS956& Chip 3& 16-11& Device 1\\
\\
JS956& Chip 3& 18-7& Device 2\\
\\
JS956& Chip 3& 18-9& Device 3\\
\\
JS956& Chip 1& 2-3& Device 4\\
\\
JS956& Chip 1& 2-9& Device 5\\
\\
JS850& Chip 1& 17-4& Device 6\\
\\
JS850& Chip 1& 17-6& Device 7\\
\\
\end{tabular}
\end{ruledtabular}
\caption{Summary of the planar Josephson junctions reported in the manuscript.}
\end{table*}

\begin{figure}[t]
    \centering
    \includegraphics[width=0.8\textwidth]{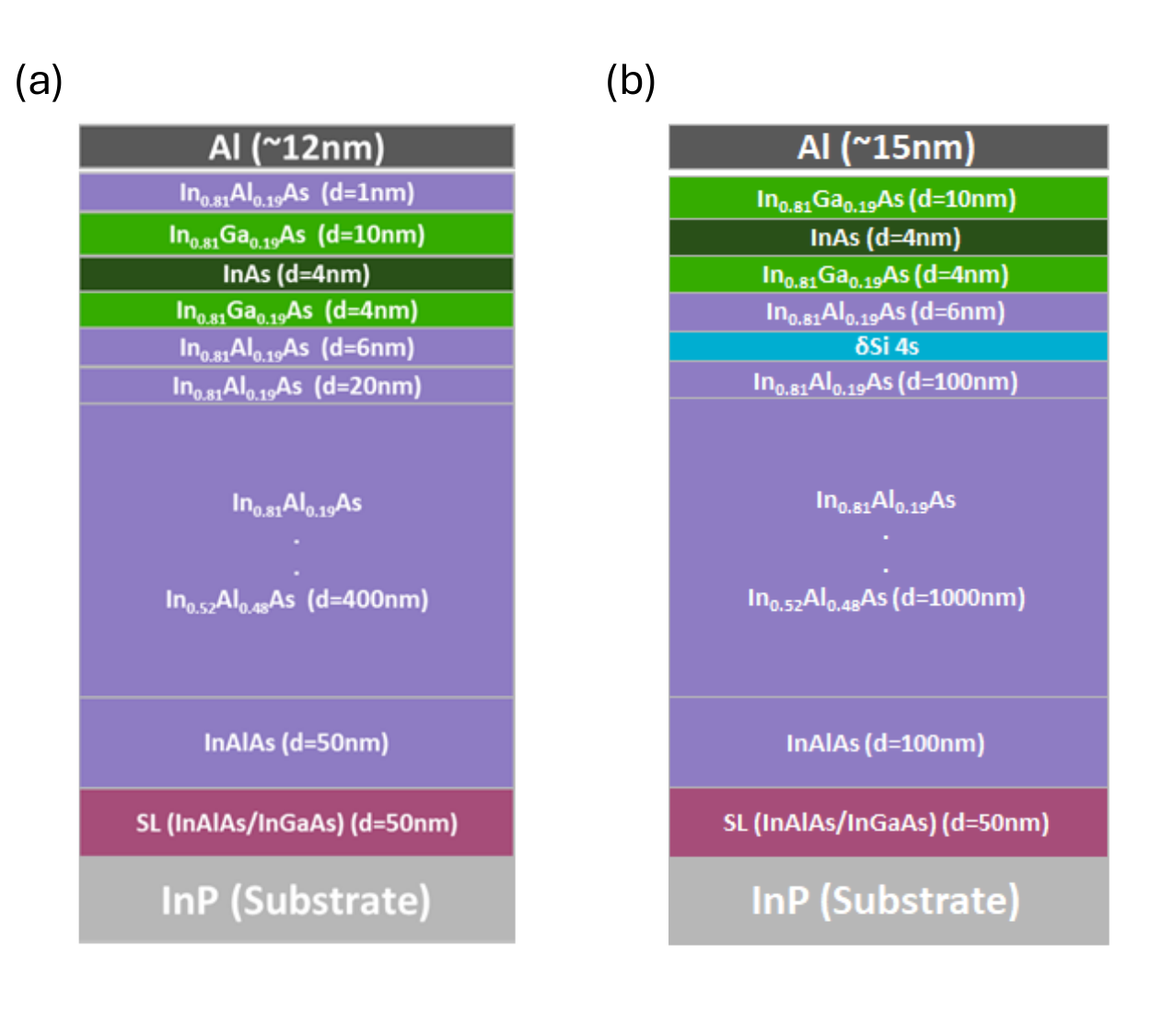}
    \caption {Layer schematic of (a) substrate JS956, (b) substrate JS850.}
    
    \label{Reentrant_Device1_MoreGates}
\end{figure}

\section{Fabrication Details}
We first make global markers on the chip using e-beam lithography (EBL) and deposit 5 nm Ti/45 nm Au. Since we have several devices on any given chip, in order to fabricate the planar junctions, we first perform EBL and etch away the Al and III-V layers around the devices to electrically isolate them from each other. Next, we perform EBL and etch Al to create a break in the Al layer which defines the junctions. This is followed by atomic layer deposition of dielectric HfO$_x$ at 80$^\circ$C using TDMAH-Hf and water vapor as precursors with a pump time of 180 seconds after each precursor pulse. We then perform EBL for top gates and deposit 10 nm Ti/390 nm Au. We use 950 PMMA A4 e-beam resist for markers, mesa etch and junction writing in EBL. For gates, we use a thicker bilayer MMA EL13/950 PMMA A4 resist. The chip is developed for 1 minute in MIBK:IPA (1:3) for 60 seconds, followed by a 60 seconds rinse in IPA. Al is etched in Transene D placed in a water bath at 50$^\circ$C. The III-V layers are etched using a 220:55:3:3 solution of DI water:1M citric acid:38\% H$_3$PO$_4$:H$_2$O$_2$ by volume. Lift-off after metal deposition is done in 50$^\circ$C acetone for one hour. 

\begin{figure}[t]
    \centering
    \includegraphics[width=0.9\textwidth]{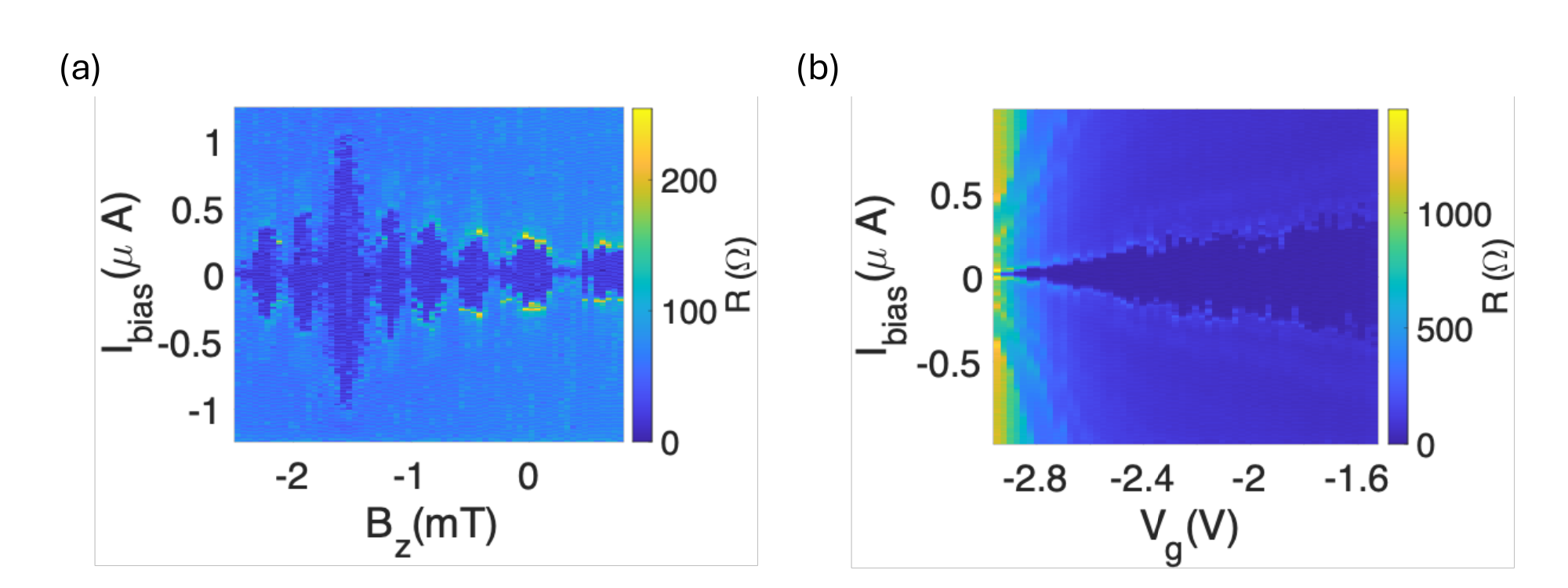}
    \caption {(a) Fraunhofer pattern, (b) Differential resistance as a function of current bias and top gate voltage for Device 1 at zero in-plane field. A lockin excitation of 10 nA was used for these two data sets.}
    
    \label{Device1_Fraunhofer_GateScan}
\end{figure}

\begin{figure}[t]
    \centering
    \includegraphics[width=0.9\textwidth]{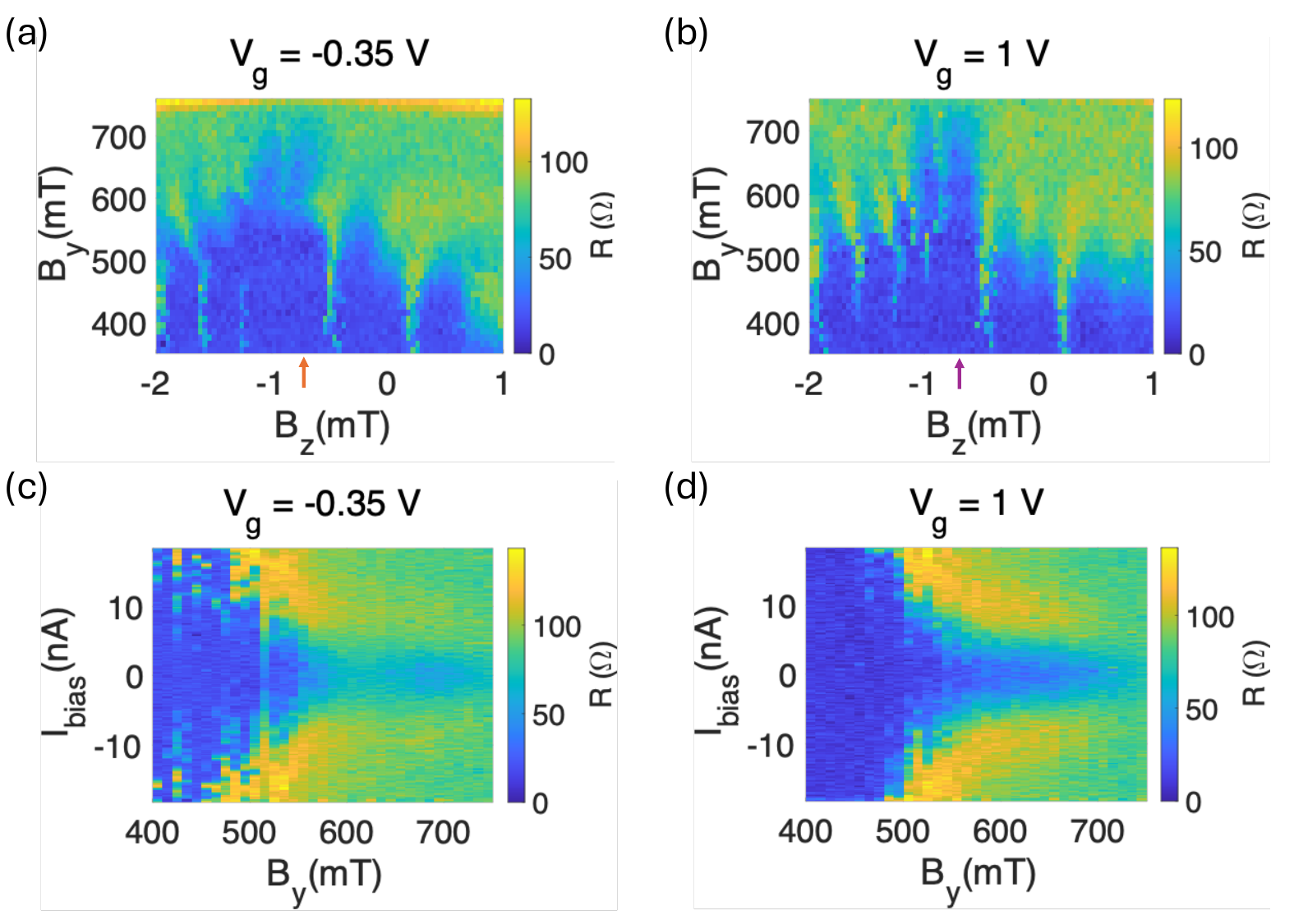}
    \caption {Supercurrent interference for Device 1 at additional gate voltage values : (a) Differential resistance as a function of the in-plane field $B_{y}$ and out-of-plane field $B_z$ at $V_g$ = -0.35 V, (b) Differential resistance as a function of the in-plane field $B_{y}$ and out-of-plane field $B_z$ at $V_g$ = 1 V, (c) Differential resistance as a function of the current bias and $B_y$ at $V_g$ = -0.35 V and $B_z$ = -0.77 mT (corresponds to orange arrow in (a)), (d) Differential resistance as a function of the current bias and $B_y$ at $V_g$ = 1 V and $B_z$ = -0.77 mT (corresponds to purple arrow in (b)). A lockin excitation of 2 nA at 43 Hz was used for these data sets.}
    
    \label{Reentrant_Device1_MoreGates}
\end{figure}

\begin{figure}[t]
    \centering
    \includegraphics[width=0.9\textwidth]{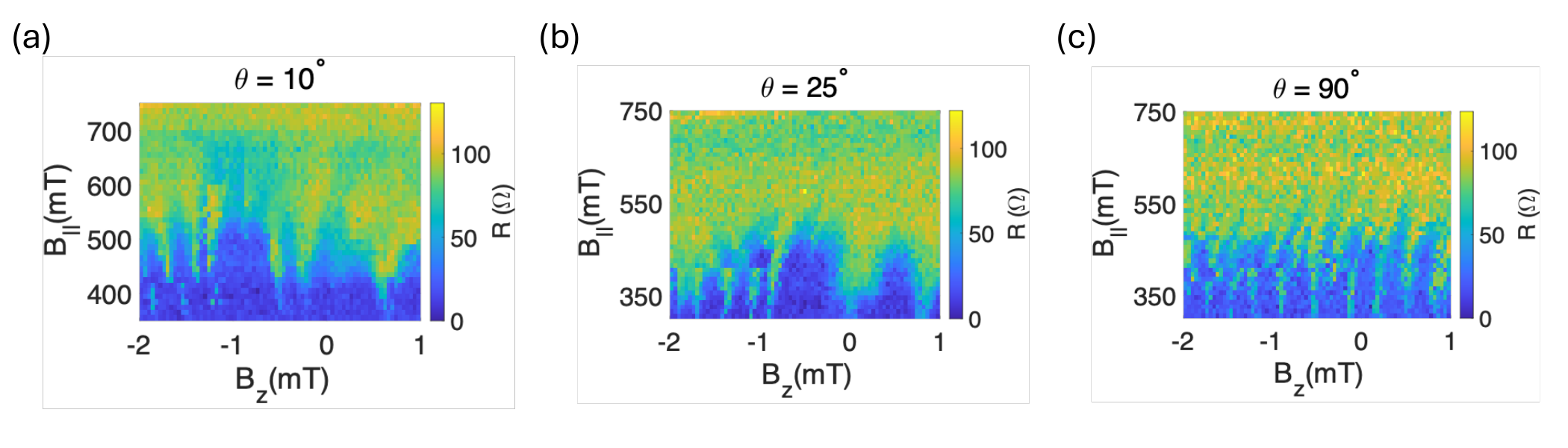}
    \caption {Supercurrent interference in-plane field angle dependence for Device 1 : Differential resistance as a function of the in-plane and out-of-plane magnetic fields $B_{||}$ and $B_z$ for $\theta$ = 10$^o$, 25$^o$ and 90$^o$ with respect to the x axis. Top gate voltage was set to 0 V. A lockin excitation of 2 nA at 43 Hz was used for these data sets.}
    
    \label{Reentrant_Device1_FieldAngleDependence}
\end{figure}

\begin{figure}[t]
    \centering
    \includegraphics[width=0.9\textwidth]{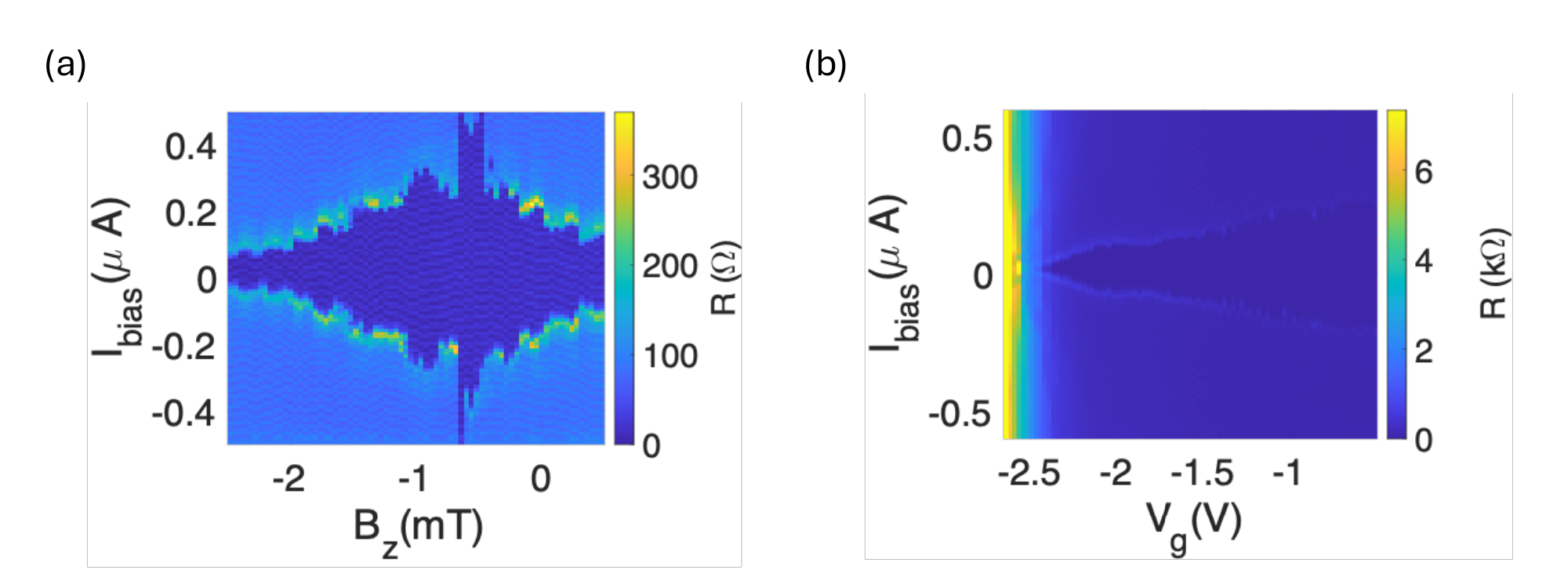}
    \caption {(a) Fraunhofer pattern, (b) Differential resistance as a function of current bias and top gate voltage for Device 2 at zero in-plane field. A lockin excitation of 10 nA was used for these two data sets.}
    
    \label{Device2_Fraunhofer_GateScan}
\end{figure}

\begin{figure}[t]
    \centering
    \includegraphics[width=0.9\textwidth]{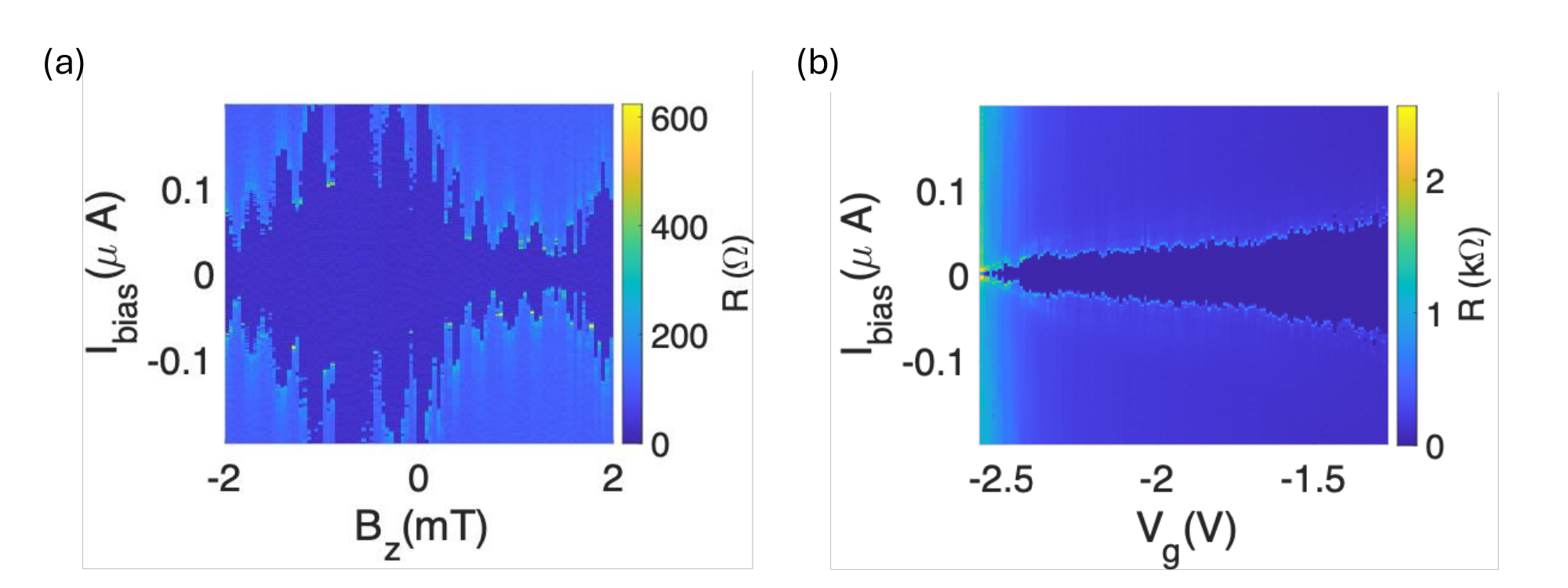}
    \caption {(a) Fraunhofer pattern, (b) Differential resistance as a function of current bias and top gate voltage for Device 3 at zero in-plane field. A lockin excitation of 10 nA was used for these two data sets..}
    
    \label{Device3_Fraunhofer_GateScan}
\end{figure}

\begin{figure}[t]
    \centering
    \includegraphics[width=0.9\textwidth]{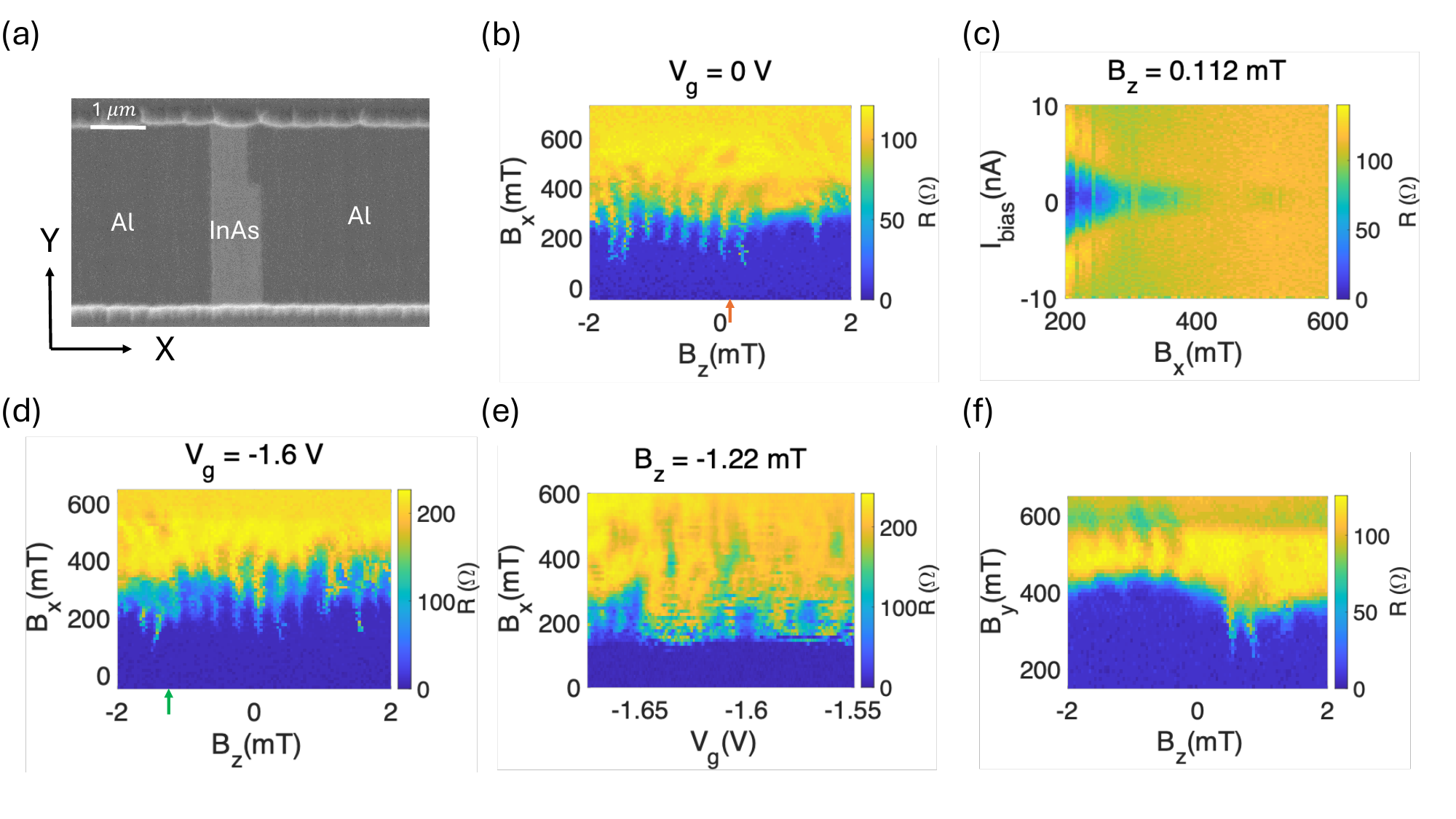}
    \caption {Supercurrent interference data for Device 3: (a) SEM Image of Device 3. The image was taken before the deposition of a top gate, (b) Differential resistance as a function of the in-plane and out-of-plane magnetic fields $B_x$ and $B_z$ at $V_g$ = 0 V. $B_x$ is applied approximately along the current flow direction, (c) Differential resistance as a function of the current bias and in-plane field $B_x$ at $B_z$ = 0.112 mT (orange arrow in (b)) and $V_g$ = 0 V, (d) Differential resistance as a function of the in-plane and out-of-plane magnetic fields $B_x$ and $B_z$ at $V_g$ = -1.6 V, (e) Differential resistance as a function of the in-plane field $B_x$ and top gate voltage $V_g$ at $B_z$ = -1.22 mT (green arrow in (c)), (f) Differential resistance as a function of the in-plane field $B_y$ and out-of-plane field $B_z$. $B_y$ is applied approximately perpendicular to the current flow direction. A lockin excitation of 1 nA was used for these data sets.}
    
    \label{Reentrant_Device3}
\end{figure}

\begin{figure}[t]
    \centering
    \includegraphics[width=0.9\textwidth]{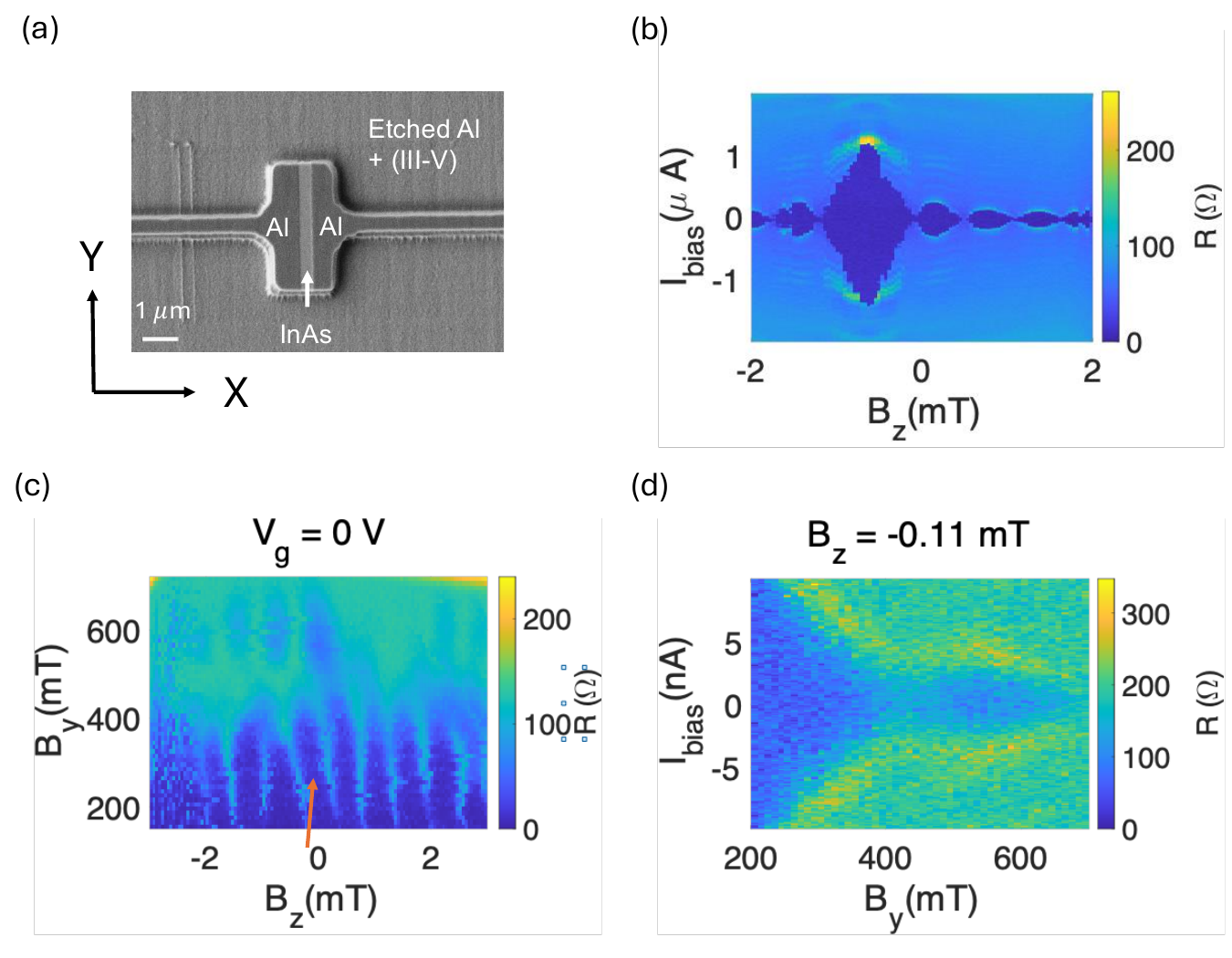}
    \caption {Supercurrent interference data for Device 4: (a) SEM Image of Device 4, (b) Fraunhofer pattern for the device for an out-of-plane field $B_z$, (c) Differential resistance as a function of the in-plane field $B_x$ and out-of-plane field $B_z$. $B_x$ is roughly perpendicular to the current-flow direction, (d) Differential resistance as a function of the current bias and $B_x$ taken along the orange arrow in (c). A lockin excitation of 1 nA was used for these data sets. This device was fabricated on wafer JS956, Chip 1 and did not have a top gate.}
    
    \label{Reentrant_Device4}
\end{figure}

\begin{figure}[t]
    \centering
    \includegraphics[width=0.9\textwidth]{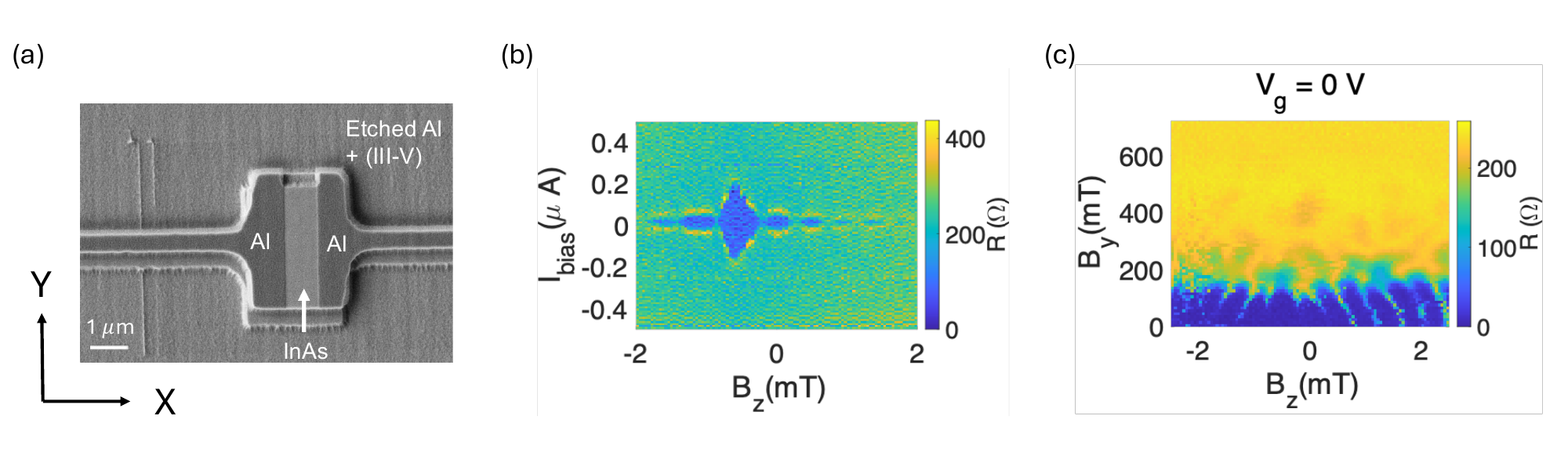}
    \caption {Supercurrent interference data for Device 5: (a) SEM Image of Device 5, (b) Fraunhofer pattern for the device for an out-of-plane field $B_z$, (c) Differential resistance as a function of the in-plane field $B_x$ and out-of-plane field $B_z$. $B_x$ is roughly perpendicular to the current-flow direction. A lockin excitation of 1 nA was used for these data sets. This device was fabricated on wafer JS956 Chip 1 and did not have a top gate.}
    
    \label{Reentrant_Device5}
\end{figure}

\begin{figure}[t]
    \centering
    \includegraphics[width=0.9\textwidth]{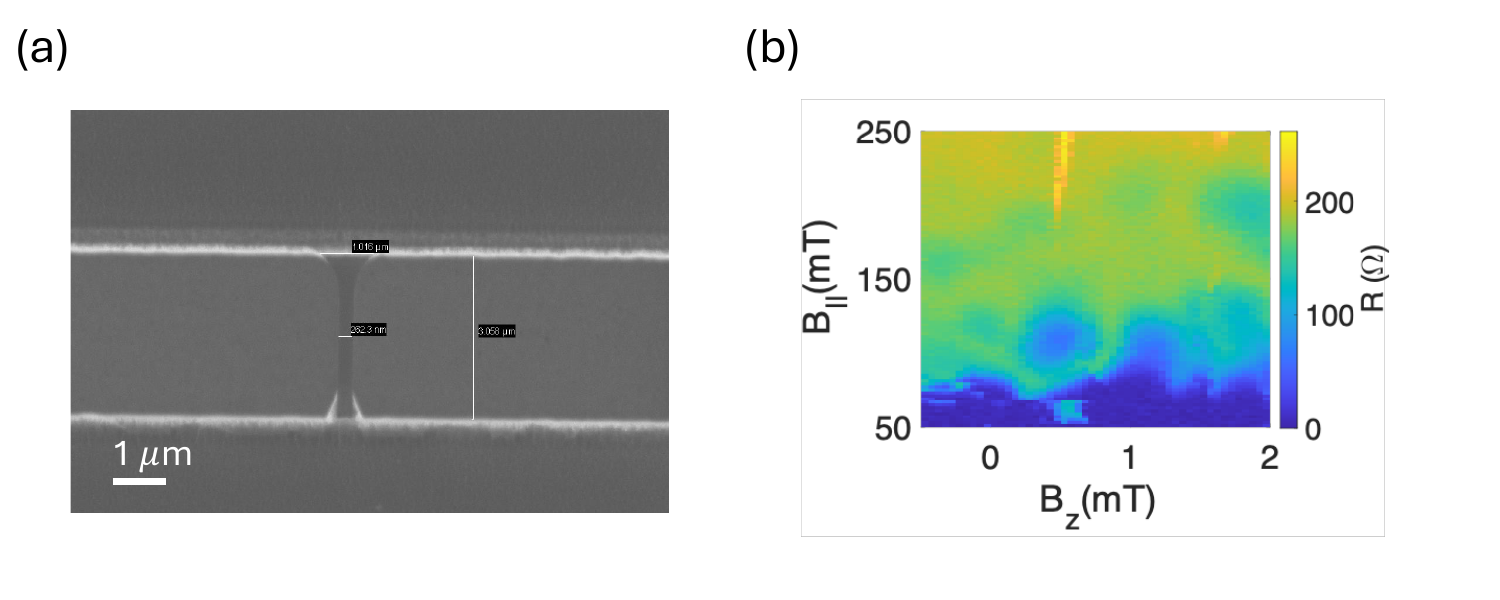}
    \caption {Supercurrent interference data for Device 6: (a) SEM Image of Device 6, (b) Differential resistance as a function of the in-plane field $B_x$ and out-of-plane field $B_z$. Here, the angle between $B_x$ and the current flow direction is not known. A lockin excitation of 1 nA at 77 Hz was used for this device. This device was fabricated on wafer JS850 Chip 1 and did not have a top gate.}
    
    \label{Reentrant_Device6}
\end{figure}

\begin{figure}[t]
    \centering
    \includegraphics[width=0.9\textwidth]{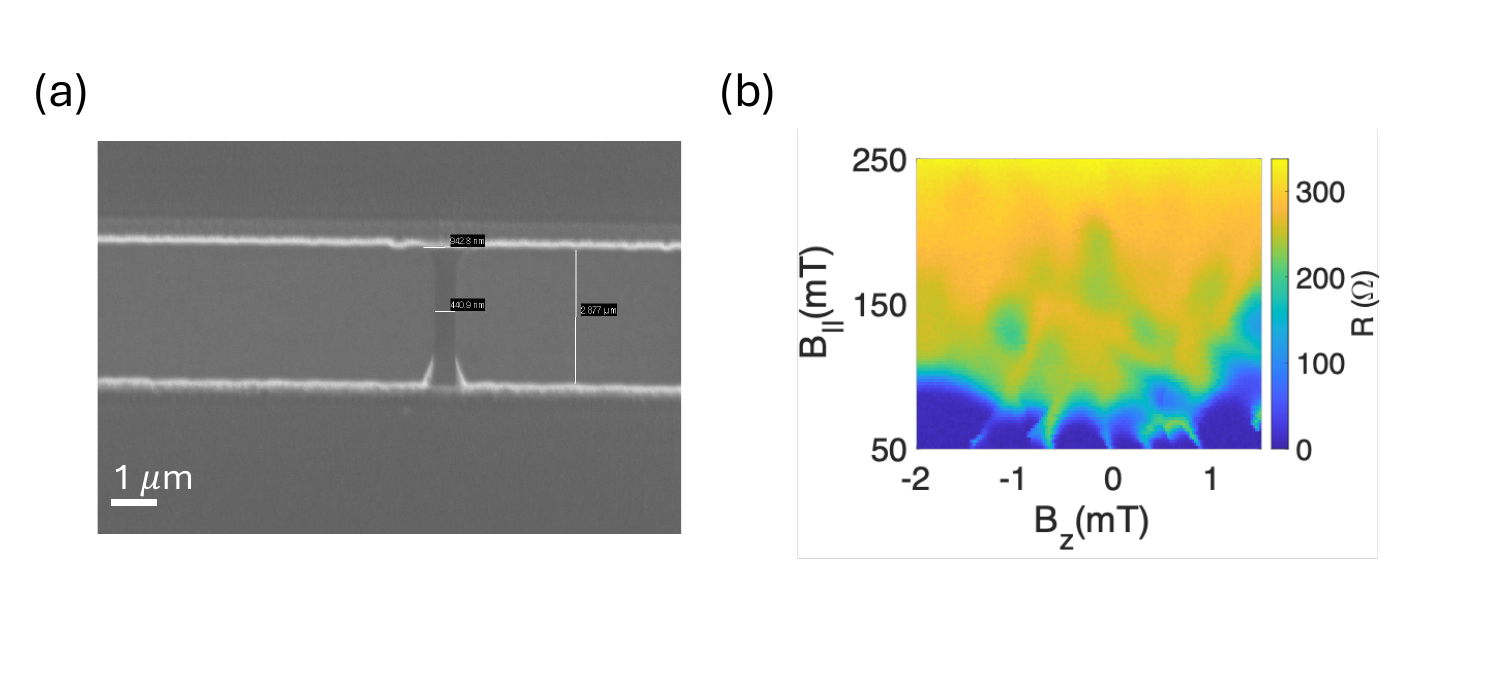}
    \caption {Supercurrent interference data for Device 7: (a) SEM Image of Device 7, (b) Differential resistance as a function of the in-plane field $B_x$ and out-of-plane field $B_z$. Here, the angle between $B_x$ and the current flow direction is not known. A lockin excitation of 1 nA at 77 Hz was used for this device. This device was fabricated on wafer JS850 Chip 1 and did not have a top gate.}
    
    \label{Reentrant_Device7}
\end{figure}

\begin{figure}[t]
    \centering
    \includegraphics[width=0.7\textwidth]{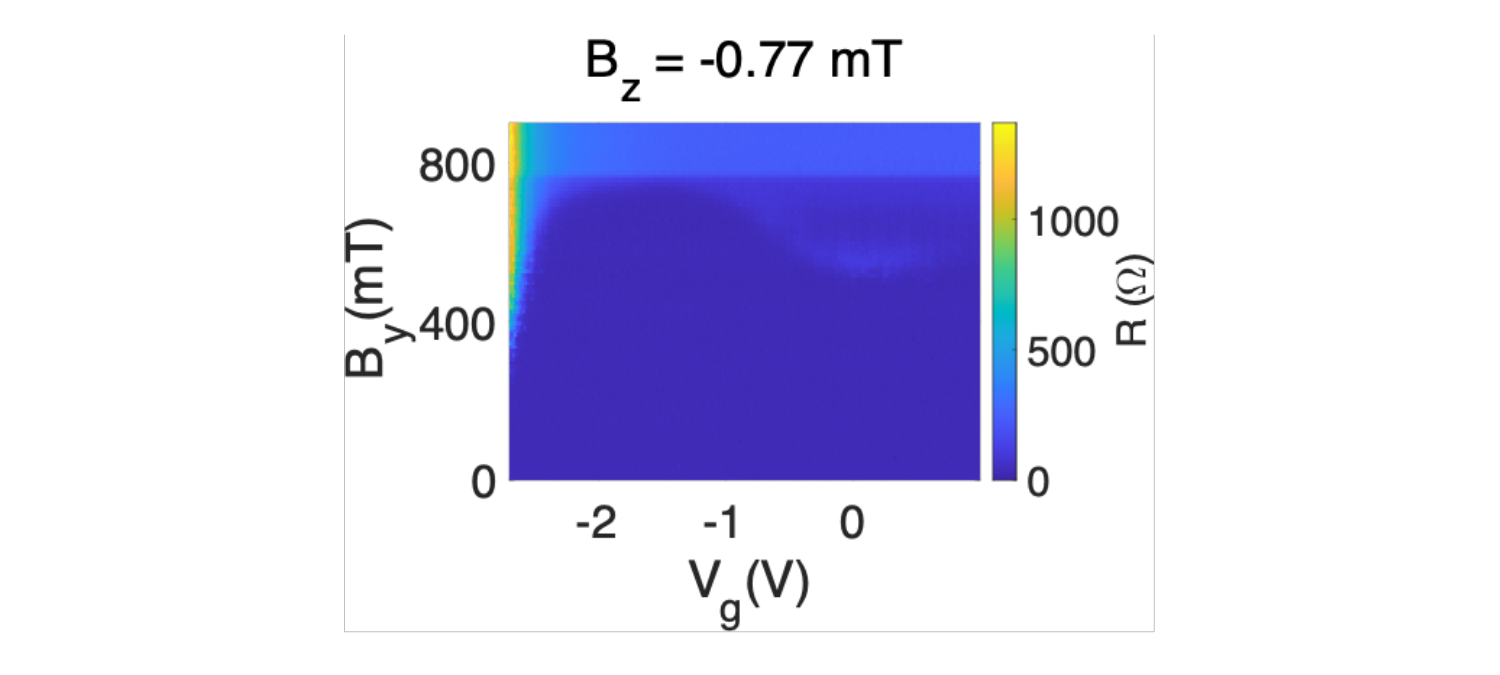}
    \caption {Raw data for Fig. \ref{Reentrant_Fig2_UniformJunctionGateDependence} (d).}
    
    \label{Fig2d_full}
\end{figure}

\begin{figure}[t]
    \centering
    \includegraphics[width=0.9\textwidth]{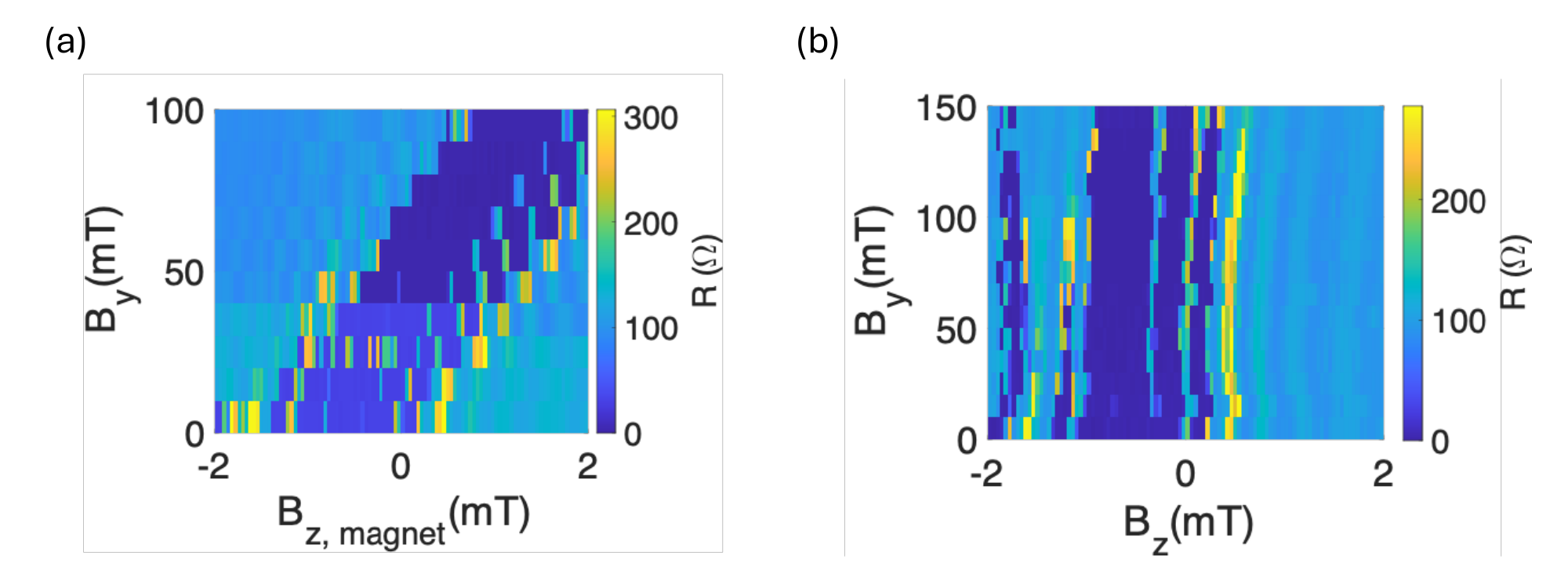}
    \caption {Device 3 : (a) Fraunhofer pattern shifting to higher z-field with increasing in-plane magnetic field $B_y$. $B_{z, magnet}$ indicates the actual field applied in the z-direction, (b) $B_y$ vs the effective z-field ($B_z$) obtained by adjusting for the slope in (a). The actual z-field at each applied in-plane field $B_y$ is adjusted in real-time. The scans were taken for a DC bias of 100 nA and a lockin excitation of 10 nA.}
    \label{ShiftedFraunhofer}
\end{figure}

\end{document}